\documentstyle[preprint,floats,prd,aps]{revtex}
\input epsf.tex
\def\DESepsf(#1 width #2){\epsfxsize=#2 \epsfbox{#1}}
\begin{document}
\preprint{\vbox {\hbox{RIKEN-AF-NP-265}}}  

\draft
\title{Intermediate psuedoscalar resonance contributions to $B\to
X_s\gamma \gamma$} \author{Mohammad R. Ahmady$^a$
\footnote{Email: ahmady@riken.go.jp} 
, Emi Kou$^b$\footnote{Email: kou@fs.cc.ocha.ac.jp} and Akio
Sugamoto$^b$\footnote{Email:sugamoto@phys.ocha.ac.jp}}
\address{
$^a$LINAC Laboratory, The Institute of Physical
and Chemical Research (RIKEN)\\ 2-1 Hirosawa, Wako, Saitama 351-01,
Japan \\
$^b$ Department of Physics, Ochanomizu University \\
1-1 Otsuka 2, Bunkyo-ku,Tokyo 112, Japan}

\date{August 1997}
\maketitle
\begin{abstract}
We calculate the decay rate $\Gamma (B\to X_s\gamma\gamma )$ via
intermediate
psuedoscalar charmonium $\eta_c$.  This process is thought to be the main
long-distance contribution which dominates the corresponding inclusive
rare B decay.  We point out that once the momentum dependence of
$\eta_c\to\gamma\gamma$ conversion strength, due to off-shellness
of the intermediate $\eta_c$, is taken into account, the rate of this
decay mode is reduced.  The change in differential decay rate in region
of spectrum immediately below $m_{\eta_c}$ is more significant.  On the
other hand, we point out that the unexpectedly
large branching ratio for $B\to X_s\eta'$ which is recently observed by
CLEO
could indicate a potentially larger long-distance contribution via $\eta'$
resonance.
\end{abstract}
%
\newpage
The rare decays of B meson are quite interesting since, among other
reasons, their measurement could point to some clues on physics beyond
Standard Model.  For this purpose, however, it is essential to isolate the
short-distance(SD) contributions from the long-distance(LD) background as
only the former is sensitive to virtual exotic particles from "new
physics".  For example, the rare dileptonic B decay $B\to
X_s\ell^+\ell^-$($\ell =\mu ,e$) in which the total decay rate is
dominated by the LD resonance contribution from intermediate $\psi$ and
$\psi'$ vector mesons, much theoretical attentions have been focused on
searching for observables and regions of various decay distributions where
one can probe SD physics without significant LD
interference\cite{ld,prd96}.  Therefore, for the purpose of extracting
reliable conclusions on parameters of Standard Model and beyond from rare
B decays, careful consideration of LD effects is of utmost importance.

In this paper, we examine the pure LD contribution to $B\to
X_s\gamma\gamma$ via intermediate psuedoscalar $\eta_c$.  This is expected
to be the main process dominating the corresponding flavor changing
neutral
current (FCNC) decay channel\cite{rrs}.  However, we show that a careful
consideration of the off-shellness of the intermediate $\eta_c$ could
reduce the contribution of this LD mode.  In fact, a model calculation of
$\eta_c$-photon-photon coupling reveals a drastic
suppression of this vertex form factor $f(q^2)$ when $q^2$, the invariant
mass of the two photons, 
is small as compared to when $q^2\approx m_{\eta_c}^2$.  Indeed, similar
suppression of $\psi$-$\gamma$ conversion strength on photon mass shell
as compared to when $\psi$ is on its mass-shell was shown to be
responsible for significant reduction of resonance to non-resonance
interference in dileptonic rare decays $B\to X_s \ell^+\ell^-$
\cite{prd96}.  Inclusion
of this form factor for $\eta_c\to \gamma\gamma$ in the
evaluation of $B\to X_s\gamma\gamma$ via $\eta_c$ is our main point in
this paper.  We also comment on the possibility that the recent CLEO
measurement of a large $B\to X_s\eta'$ decay rate could indicate a
potentially larger LD contribution to $B\to X_s\gamma\gamma$ via $\eta'$.

We start by writing the effective Hamiltonian for LD $B\to
X_s\gamma\gamma$ via $\eta_c$ which is approximated by quark level decay
$b\to s\eta_c\to s\gamma\gamma$.  The $\eta_c\to\gamma\gamma$ transition
is modeled by a triangle quark loop (Fig. 1) where the coupling
of the psuedoscalar meson $\eta_c$ to charm quarks is taken to be a
constant:
\begin{equation}
T^{\mu\nu}(\eta_c\to\gamma\gamma )=
Nf(q^2)\epsilon^{\mu\nu\alpha\beta}p_{1\alpha}p_{2\beta} \;\; ,
\end{equation}
where $p_1$ and $p_2$ are the four-momenta of the photons and $q=p_1
+p_2$.  The form factor $f(q^2)$ is obtained from the quark loop
calculation:
\begin{equation}
f(q^2)
\begin{array}[t]{l}
\displaystyle = \int^1_0dx\int_0^{1-x}dy\frac{1}{m_c^2-q^2xy}  \\
=\left  \{ { 
\begin{array}{l}
\displaystyle
\frac{-2}{q^2}arcsin^2\sqrt{\frac{q^2}{4m_c^2}}\;\;\; 0\le q^2\le 4m_c^2
\\
\displaystyle\frac{2}{q^2}{\left [ Ln\left
(\sqrt{\frac{q^2}{4m_c^2}}+\sqrt{\frac{q^2}{4m_c^2}-1}\right
)-\frac{I\pi}{2}\right ]}^2 \;\;\; 4m_c^2\le q^2
\end{array}}
\right .
\;\; ,
\end{array}
\end{equation}
where $m_c$ is the charm quark mass.  The constants are all swept into the
factor $N$ which can be obtained using the requirement that for
$q^2=m_{\eta_c}^2$ eqn. (1) should yield the experimentally measured decay
rate $\Gamma (\eta_c\to\gamma\gamma )$.  Consequently, we obtain the
following form for the $\eta_c$-$\gamma\gamma$ transition amplitude:
\begin{equation}
\displaystyle
A(\eta_c\to\gamma\gamma)= \frac{16i\sqrt{m_{\eta_c}\Gamma
(\eta_c\to\gamma\gamma
)}}{\pi^{3/2}}f(q^2)\epsilon^{\mu\nu\alpha\beta}\epsilon_\mu
(p_1)\epsilon_\nu (p_2)p_{1\alpha}p_{2\beta}\;\; .
\end{equation}
$\epsilon (p_i)$ is the polarization of the photon with momentum $p_i$.
Throughout this paper, we assume weak binding for charmonium and therefore
$m_{\eta_c}\approx 2m_c$ is used in our calculations.

Neglecting penguin operators, the relevant effective Hamiltonian for $B\to
X_s\eta_c$ can be written as:
\begin{equation}
H_{eff}=\displaystyle \frac{G_F}{\sqrt{2}}V_{cb}^*V_{cs}
\begin{array}[t]{l} \left [ \left  ( C_2(\mu ) + \frac{1}{3}C_1(\mu ) 
\right ) 
                 {\bar s}^i\gamma^\mu (1-\gamma_5)b^i
                 {\bar c}^j\gamma_\mu (1-\gamma_5)c^j \right .\\  
                 \displaystyle\;\;  + \left . C_1(\mu ){\bar
s}^i\gamma^\mu 
                 (1-\gamma_5)T^a_{in}b^n{\bar c}^j\gamma_
                 \mu (1-\gamma_5)T^a_{jm}c^m \right ] +H.C. \; ,
\end{array}
\end{equation}
where $i$ and $j$ are color indices, $T^a$ ($a=1..8$) are generators of
$SU(3)_{\rm color}$, and $C_1(\mu )$ and $C_2(\mu )$ are QCD 
improved Wilson coefficients.  Consequently, assuming factorization, the
matrix element for the underlying quark level decay can be simplified as
follows:
\begin{equation}
<\eta_c s|H_{eff}|b>  
\begin{array}[t]{l}
\displaystyle =\frac{G_F}{\sqrt{2}}V_{cb}^*V_{cs}
\left  ( C_2(\mu ) + \frac{1}{3}C_1(\mu ) \right ) f_{\eta_c}\bar
s\gamma^\mu (1-\gamma_5)bq_\mu \\
\displaystyle =\frac{G_F}{\sqrt{2}}V_{cb}^*V_{cs}
\left  ( C_2(\mu ) + \frac{1}{3}C_1(\mu ) \right ) f_{\eta_c}
\left [ -m_s\bar s(1-\gamma_5)b+m_b\bar s(1+\gamma_5)b\right ]\;\; .
\end{array}
\end{equation}
In writing (5), the following definition for psuedoscalar decay constant
has been utilized:
\begin{equation}
<0|\bar c\gamma^\mu\gamma_5c|\eta_c(q)>=f_{\eta_c}q^\mu\;\; .
\end{equation}
Using (3) and (5), the amplitude for $B\to X_s\gamma\gamma$ via $\eta_c$
can be expressed as:
\begin{equation}
A^{LD(\eta_c)}(B\to X_s\gamma\gamma )=
\begin{array}[t]{l}
\displaystyle C f(q^2)\bar s(-m_s(1-\gamma_5)+m_b(1+\gamma_5))b \\
\displaystyle \times
\frac{1}{q^2-m_{\eta_c}^2+im_{\eta_c}\Gamma_{\eta_c}}\epsilon^{\mu\nu\alpha\beta}\epsilon_\mu
(p_1)\epsilon_\nu (p_2)p_{1\alpha}p_{2\beta}\;\; ,
\end{array}
\end{equation}
where 
\begin{equation}
C=
\displaystyle 
\frac{16iG_F}{\sqrt{2}\pi^{3/2}}\sqrt{m_{\eta_c}\Gamma
(\eta_c\to\gamma\gamma )}V_{cb}^*V_{cs}
\left  ( C_2(\mu ) + \frac{1}{3}C_1(\mu ) \right ) f_{\eta_c}\;\; ,
\end{equation}
and $\Gamma_{\eta_c}$ is the total decay width of $\eta_c$.  It is then
straightforward to calculate the differential decay rate from (7):
\begin{equation}
\frac{d\Gamma^{LD(\eta_c )}(B\to X_s\gamma\gamma )}{ds}=
\begin{array}[t]{l}
\displaystyle \frac{m_b}{512\pi^3}{\vert C\vert}^2{\vert \bar
f(s)\vert}^2\frac{s^2}{{(s-y)}^2+y\frac{\Gamma^2_{\eta_c}}{m_b^2}}\\
\displaystyle \left [ {(1-x)}^2-s(1+x)\right ](1-x) \;\; ,
\end{array}
\end{equation}
where
\begin{equation}
{\vert\bar f(s)\vert}^2=\displaystyle \frac{4}{s^2} \left \{ 
\begin{array}{l}
\displaystyle
arcsin^4\sqrt{\frac{s}{y}}\;\;\; 0\le s\le y \\
\displaystyle {\left [ Ln^2\left
(\sqrt{\frac{s}{y}}+\sqrt{\frac{s}{y}-1}\right )+\frac{\pi^2}{4}\right
]}^2 \;\;\; y\le s
\end{array}
\right .
\;\; .
\end{equation}
The dimensionless quantities $s$, $x$ and $y$ are $q^2/m_b^2$,
$m_s^2/m_b^2$ and $m_{\eta_c}^2/m_b^2$, respectively.  In Fig. 2, the form
factor ${\vert \bar f(s)\vert}^2$ normalized to its value on $\eta_c$
mass-shell ${\vert \bar f(y)\vert}^2$ is depicted.  We notice that for
values of $q^2$ immediately below $m_{\eta_c}^2$, ${\vert
f(q^2)\vert}^2$ decreases steeply from ${\vert
f(q^2=m_{\eta_c}^2)\vert}^2$.  In fact,  ${\vert f^2(0)\vert}^2/{\vert
f^2(q^2=m^2_{\eta_c})\vert}^2\approx 0.16$ which indicates that the
momentum dependence of this form factor is quite significant.  As we
mentioned before, a similar mechanism for intermediate vector mesons
$\psi$ and $\psi'$ is believed to suppress the LD contributions to rare
decay $B\to X_s\gamma$ \cite{bsg} and significantly reduces the resonance
to nonresonance interference in dileptonic rare B decays $B\to
X_s\ell^+\ell^-$\cite{prd96}.

In our numerical calculations $m_b$, $m_s$ and $f_{\eta_c}$ are taken as
$4.8$, $0.5$ and $0.48$ GeV, respectively, and $C_1(\mu )+1/3C_2(\mu
)=0.155$ for $\mu\approx m_b$ is adopted from next-to-leading order
calculation\cite{buras}.  Using
\begin{equation}
\displaystyle
\frac{G_f^2m_b^5{\vert V_{cb}\vert}^2}{192\pi^3 \Gamma_B} \approx 0.2\;\;
,
\end{equation}
from semileptonic $B$ decay $B\to X_c\ell\bar{\nu_\ell}$, we obtain
\begin{equation}
BR^{LD(\eta_c )}(B\to X_s\gamma\gamma )=9.1\times 10^{-7}\;\; ,
\end{equation}
which is of the same order of magnitude as the estimated $(2-8)\times
10^{-7}$ SD contributions\cite{rrs}.  At this point, we would like to
remark that the total LD branching ratio (via $\eta_c$) is dominated by
the peak at $s=m^2_{\eta_c}/m^2_b$.  However, had we inserted the constant
form factor $f(q^2=m_{\eta_c}^2)$ in (9) rather than $f(q^2)$, a larger
branching ratio $BR^{LD(\eta_c )}(B\to X_s\gamma\gamma )=10.1\times
10^{-7}$ would have been resulted.  In fig. 3, the invariant mass
distribution of the decay rate has been shown (eqn. (9)) and for
comparison, the case where $f(q^2=m^2_{\eta_c})$ replaces $f(q^2)$ in (9)
is depicted as well.  We observe that due to the significant decrease in
$\eta_c$-photon-photon coupling strength for $q^2$ values immediately
below $m_{\eta_c}^2$, the LD differential decay rate in this region is
further reduced.  For example, even at $s=0.2$ which is not too close
to the resonance, the reduction factor is around $1/3$.  This means that
a wider range of invariant mass spectrum
below $\eta_c$ resonance $0\le s\le 0.39$ is available  for probing SD
physics.

Now, we turn to another potentially large source of LD contribution to
$B\to X_s\gamma\gamma$.  Recently, CLEO discovered an unexpectedly large
branching ratio for $B\to X_s\eta'$\cite{cleo}
\begin{equation}
BR(B\to X_s\eta'\;\;\;\; 2.2\le E_{\eta'}\le 2.7 {\rm GeV})
=(7.5\pm 1.5\pm 1.1)\times 10^{-4} \;\; .
\end{equation}
This decay mode can contribute to LD $B\to X_s\gamma\gamma$ by subsequent
$\eta'$ decay to two photons.  To make a rough estimate of this
contribution, we compare the relevant branching ratios for $\eta'$ and
$\eta_c$ cases:
\begin{equation}
\displaystyle
\frac{BR(B\to X_s\eta' )BR(\eta'\to\gamma\gamma )}{BR(B\to
X_s\eta_c)BR(\eta_c\to\gamma\gamma )}\approx 6 \;\; ,
\end{equation}
where $BR(B\to X_s\eta_c )\approx 8.7\times 10^{-3}$ has been
used\cite{ak}.  This could be an indication that the LD $B\to
X_s\gamma\gamma$ via $\eta'$ may surpass that of $\eta_c$.  However, to
calculate this contribution more accurately, we need to know the mechanism
of such a large $\eta'$ production in nonleptonic $B$ decay.
Unfortunately, so far a clear process for $B\to X_s\eta'$ has not been
established yet and there are various reservations with regard to
suggested mechanisms\cite{etap}.  In any case, a large peak at low
$s=m_{\eta'}^2/m_b^2\approx 0.04$ is expected.

In conclusion, we calculated the LD decay process $B\to X_s\gamma\gamma$
via $\eta_c$ taking into account the momentum dependence of
$\eta_c$-photon-photon vertex for off mass-shell $\eta_c$.  We indicated
that a wider range of the invariant mass spectrum below $\eta_c$ resonance
could be available for probing SD physics.  However, the decay mode via
$\eta'$ could be a larger source of LD background to $B\to
X_s\gamma\gamma$ rare decay which ought to be investigated once the
mechanism for $B\to X_s\eta'$ is established.

\vskip 1.0cm
{\bf \large Acknowledgement}

The authors would like to thank T. Morozumi for useful 
discussions.  M. A. acknowledges support from the Science and Technology
Agency of Japan under an STA fellowship.

\newpage
{\center \bf \huge Figure Captions} 
\vskip 3.0cm
\noindent
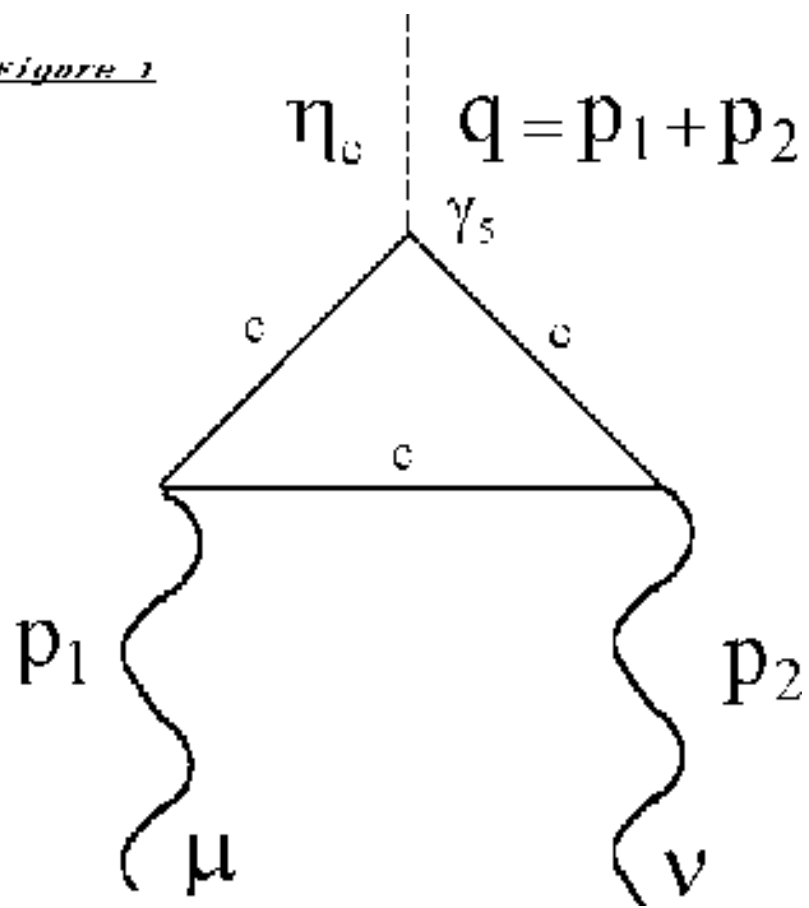
{\bf Figure 1}: The triangle quark loop diagram for $\eta_c$-photon-photon
coupling. \\
\vskip 0.5cm
\noindent
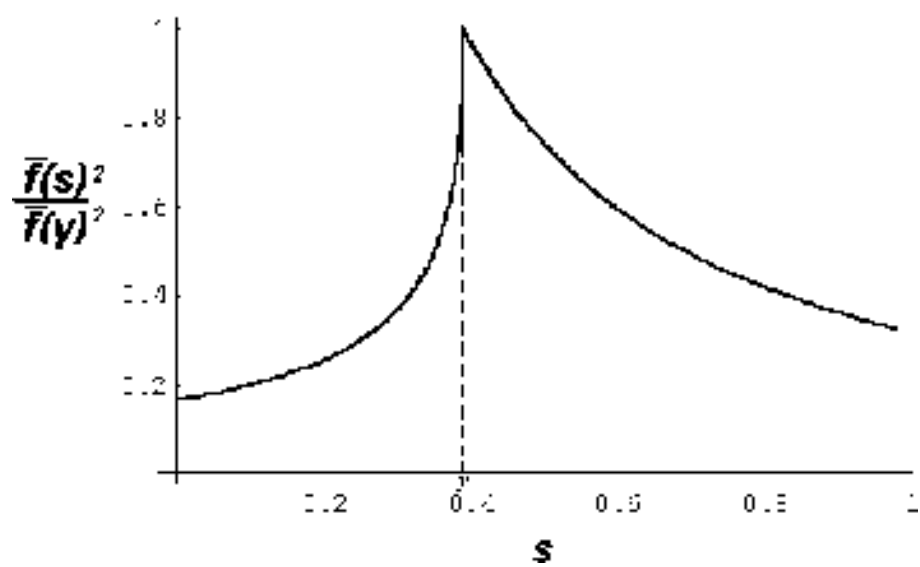
{\bf Figure 2}: The variation of the form factor ${\vert\bar
f(s=q^2/m_b^2)\vert}^2$ normalized to ${\vert\bar
f(y=m_{\eta_c}^2/m_b^2)\vert}^2$ as a function of $s$. \\
\vskip 0.5cm
\noindent
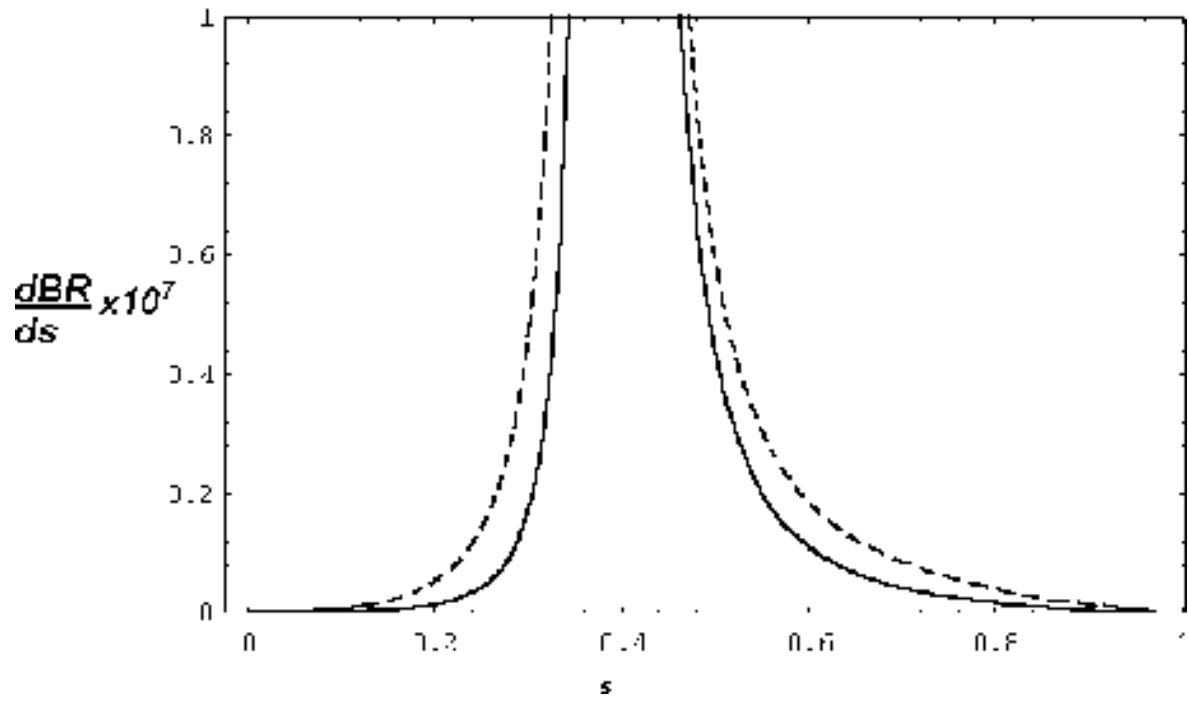
{\bf Figure 3}: The invariant mass spectrum of the two photons in $B\to
X_s\gamma\gamma$ decay for momentum dependent (solid line) and constant
(dashed line) $\eta_c$-photon-photon coupling.  The differential branching
ratio is in units of $10^{-7}$.

\end{document}